\documentclass[a4paper,11pt]{article}
\usepackage{pos}

\title{$K\to\pi\pi$ decay matrix elements at the physical point with periodic boundary conditions}
\ShortTitle{$K\to\pi\pi$ with periodic boundaries}

\author*[a]{Masaaki~Tomii}
\author[a,b]{Thomas~Blum}
\author[c]{Daniel~Hoying}
\author[b,d]{Taku~Izubuchi}
\author[a,b]{Luchang~Jin}
\author[d]{Chulwoo~Jung}
\author[d]{Amarjit~Soni}

\affiliation[a]{Physics Department, University of Connecticut, Storrs, CT 06269, USA}
\affiliation[b]{RIKEN-BNL Research Center, Brookhaven National Laboratory, Upton, NY 11973, USA}
\affiliation[c]{Department of Computational Mathematics, Science and Engineering, and Department of Physics and Astronomy, Michigan State University, East Lansing, MI, 48824, USA}
\affiliation[d]{Physics Department, Brookhaven National Laboratory, Upton, NY 11973, USA}

\emailAdd{masaaki.tomii.at.uconn.edu}

\abstract{
We calculate $K\to\pi\pi$ matrix elements using periodic boundary conditions as an independent calculation from our previous study with G-parity boundary conditions. We present our preliminary results for $K\to\pi\pi$ three-point functions and matrix elements on a $24^3, a^{-1} = 1$~GeV, $2+1$-flavor M\"obius DWF ensemble at physical pion and kaon masses generated by the RBC and UKQCD collaborations and discuss the prospect for high-precision computation of $\varepsilon'$ with periodic boundary conditions.
}

\FullConference{%
 The 38th International Symposium on Lattice Field Theory, LATTICE2021
  26th-30th July, 2021
  Zoom/Gather@Massachusetts Institute of Technology
}

\begin{document}
\maketitle

\section{Introduction}
\label{sec:intro}

Direct CP violation in $K\to\pi\pi$ decays, which is parametrized
by $\varepsilon'$,
attracts close attention from particle phenomenologists.
Since it is highly suppressed in the Standard Model, it is expected
to be a rich prove for searching new physics beyond the
Standard Model.

With various improvements to our first result for $\varepsilon'$~\cite{RBC:2015gro},
we have completed our calculation on a single lattice
ensemble with G-parity boundary conditions~\cite{RBC:2020kdj}.
Our result is ${\rm Re}~(\varepsilon'/\varepsilon)_{\rm SM} = 21.7(2.6)(6.2)(5.0)\times10^{-4}$,
where $\varepsilon$ is the measure of indirect CP violation and the first two errors
represent statistical and systematic in our lattice calculation. 
The third error corresponds to omitted isospin breaking
and QED corrections.
While this result is consistent with the average of the experimental
results ${\rm Re}~(\varepsilon'/\varepsilon)_{\rm exp} = 16.6(2.3)\times10^{-4}$~\cite{NA48:2002tmj,KTeV:2010sng},
more independent calculations with better precision would be desired.

The major sources of systematic error in our previous calculation are
1.~truncation error of perturbative matching of
the Wilson coefficients between four- and three-flavor theories at the
charm threshold, 2.~finite lattice cutoff effects and 3.~isospin breaking
and QED corrections.
For the matching of the Wilson coefficients, we are attempting to 
implement a nonperturbative procedure~\cite{Tomii:2020smd}
instead of the perturbative approach.

The other two sources of systematic uncertainty listed above motivate us
to calculate the $K\to\pi\pi$ matrix elements with periodic boundary
conditions.
For finite lattice cutoff effects, while we are generating finer G-parity
ensembles to take the continuum limit within the calculation with
G-parity boundary conditions, we can perform an independent
calculation with periodic boundary conditions with the ensembles with
various lattice spacings that we have already generated.
Also, it is more straightforward for periodic boundary conditions to introduce
isospin breaking and QED effects than G-parity boundary conditions.
Thus our study would give us the prospect of obtaining $\varepsilon'$ more
precisely.

It is challenging to extract the on-shell kinematics of $K\to\pi\pi$ from
euclidean three-point functions with periodic boundary conditions
because of the presence of the off-shell two-pion state with the
energy $E_{\pi\pi} \simeq 2m_\pi$, which is smaller than that of the on-shell
kinematics $E_{\pi\pi} = m_K$.
It also turned out that the contamination from heavier two-pion states
are significant~\cite{RBC:2020kdj,RBC:2021acc}.
Therefore we seriously perform state decomposition by solving the generalized
eigenvalue problem (GEVP)~\cite{Blossier:2009kd,Bulava:2011yz}
to well extract the on-shell kinematics of $K\to\pi\pi$ decays.

In this article we give our preliminary results for the $K\to\pi\pi$ three-point
functions and matrix elements calculated on the $24^3\times64$ lattice
ensemble with $2+1$-flavor M\"obius domain-wall fermions near the
physical pion and kaon masses and the lattice cutoff $a^{-1} \simeq 1$~GeV.
The measurements are done with 258 configurations.
We use all-to-all quark propagators with 2,000 low modes from the
Lanczos algorithm for the light quark and spin-color-time diluted
CG inversions with random noise vectors for both the light and
strange quarks.

\section{Correlation functions}
\label{sec:correlators}

In order to obtain the $K\to\pi\pi$ matrix elements, we need to
calculate two-point functions of interpolation operators as well as
three-point functions of the interpolation operators and the $\Delta S = 1$
four-quark effective operators.
The following subsections explain these correlation functions.

\subsection{Two-point function of kaon operator}
\label{sec:kaon2pt}

The simplest correlation function is the two-point function of a kaon
interpolation operator 
\begin{equation}
C^K(t) = \left\langle O_K(t)^\dag O_K(0)\right\rangle.
\end{equation}
We perform the exponential smearing for the kaon interpolation operator with
the radius of $2$ in lattice units.

\subsection{Two-point functions of two-pion operators}
\label{sec:pipi2pt}

We employ multiple two-pion interpolation operators 
\begin{equation}
O_{\pi\pi,\alpha}(t) \in \{\pi\pi(000), \pi\pi(001), \pi\pi(011), \pi\pi(111), \sigma\},
\label{eq:pipiOps}
\end{equation}
where $\pi\pi(abc)$ with $a,b,c = 0,1$ is a product of two pion interpolation operators
with the momenta indicated by $(abc)$.
The number of the 1's in $abc$ indicates the number of directions to which each
pion interpolation operator has one unit of momentum $\pm 2\pi/L$.
One of the pion interpolation operators has the opposite momentum to the other's,
i.e. we employ the center of mass frame.
We use neutral isospin-definite two-pion operators, which can be
constructed by an appropriate linear combination of $\pi^+\pi^-$, $\pi^0\pi^0$ and
$\pi^-\pi^+$.
For the $I = 0$ channel, we also employ $\sigma$ as a two-pion interpolation operator.
We perform the exponential smearing for the pion and sigma operators with
the radius of $1.5$ in lattice units.

We calculate two-point functions of these two-pion operators
\begin{equation}
C^{\pi\pi}_{\alpha\beta}(t) 
= \left\langle O_{\pi\pi,\alpha}(t)^\dag O_{\pi\pi,\beta}(0) \right\rangle
- \left\langle O_{\pi\pi,\alpha}^\dag \right\rangle \left\langle O_{\pi\pi,\beta} \right\rangle,
\end{equation}
where the second term on the right hand side is the vacuum subtraction term
needed for the $I=0$ channel.
In order to extract the signal from a certain state, we solve the
generalized eigenvalue problem (GEVP)~\cite{Blossier:2009kd}
\begin{equation}
\sum_\beta C_{\alpha\beta}^{\pi\pi}(t) v_{n,\beta}(t,t_0)
= \lambda_n(t,t_0)\sum_\beta C_{\alpha\beta}^{\pi\pi}(t_0) v_{n,\beta}(t,t_0),
\end{equation}
where $\lambda_n(t,t_0)$ is an eigenvalue of the GEVP and
$v_{n,\beta}(t,t_0)$ is the corresponding eigenvector.
We use $t_0 = t-1$ in this work.
Preliminary results for our GEVP analysis are presented in Ref.~\cite{Hoying:2020ghg}.

\subsection{$K\to\pi\pi$ three-point functions}
\label{sec:3pt}

\begin{figure}[tbp]
\begin{center}
\begin{tabular}{c}
\includegraphics[width=60mm]{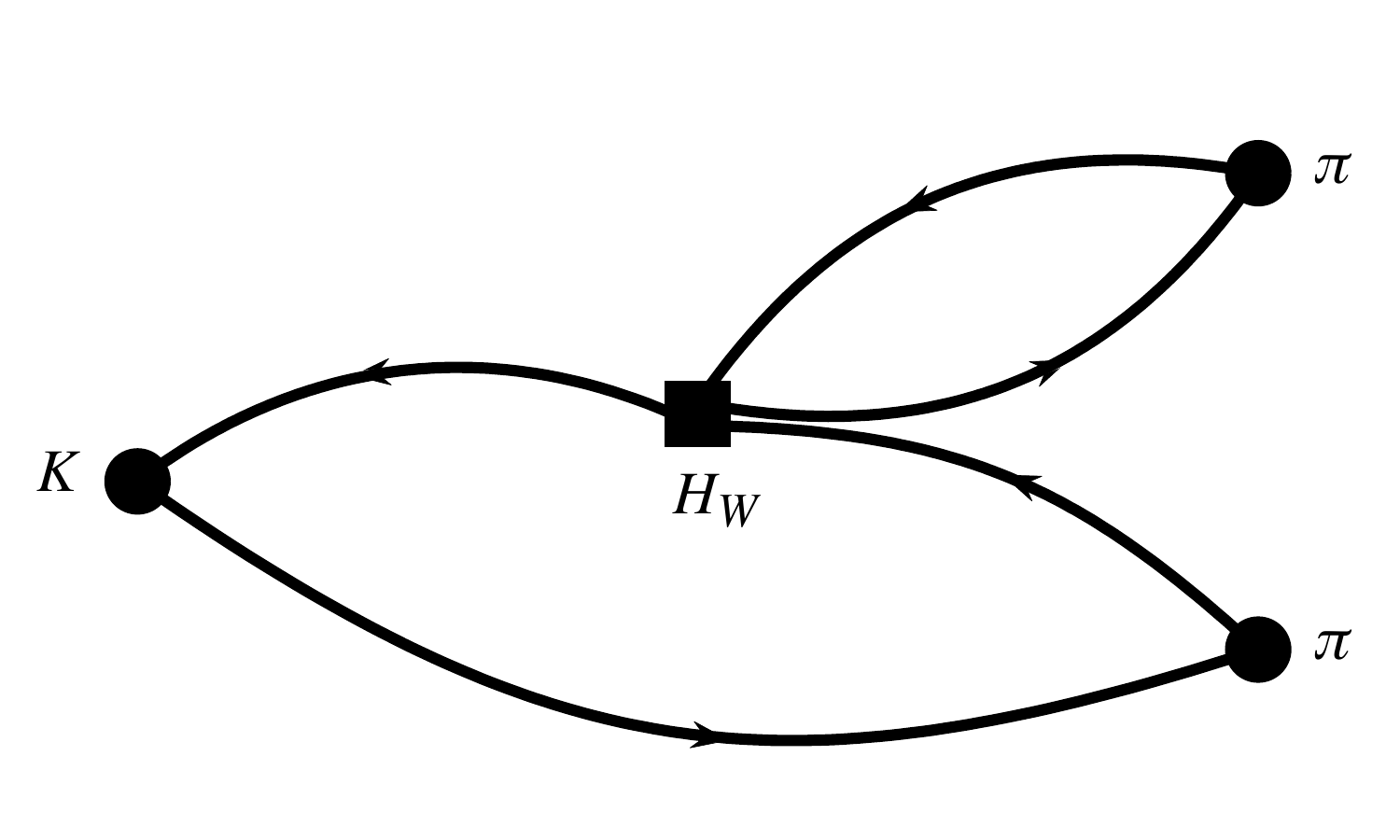}
\\
{\it type1}
\end{tabular}
\hspace{10mm}
\begin{tabular}{c}
\includegraphics[width=60mm]{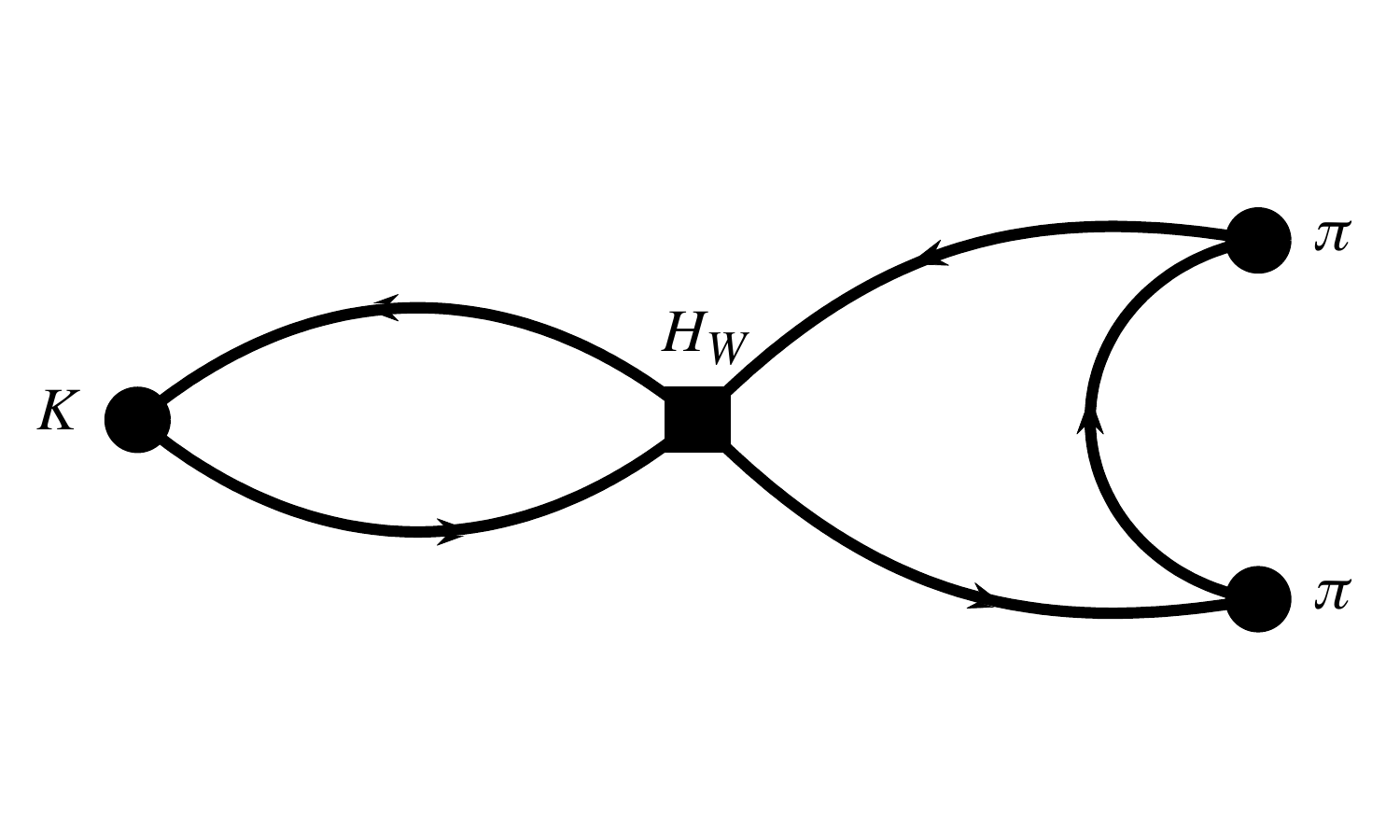}
\\
{\it type2}
\end{tabular}
\\[3mm]
\begin{tabular}{c}
\includegraphics[width=60mm]{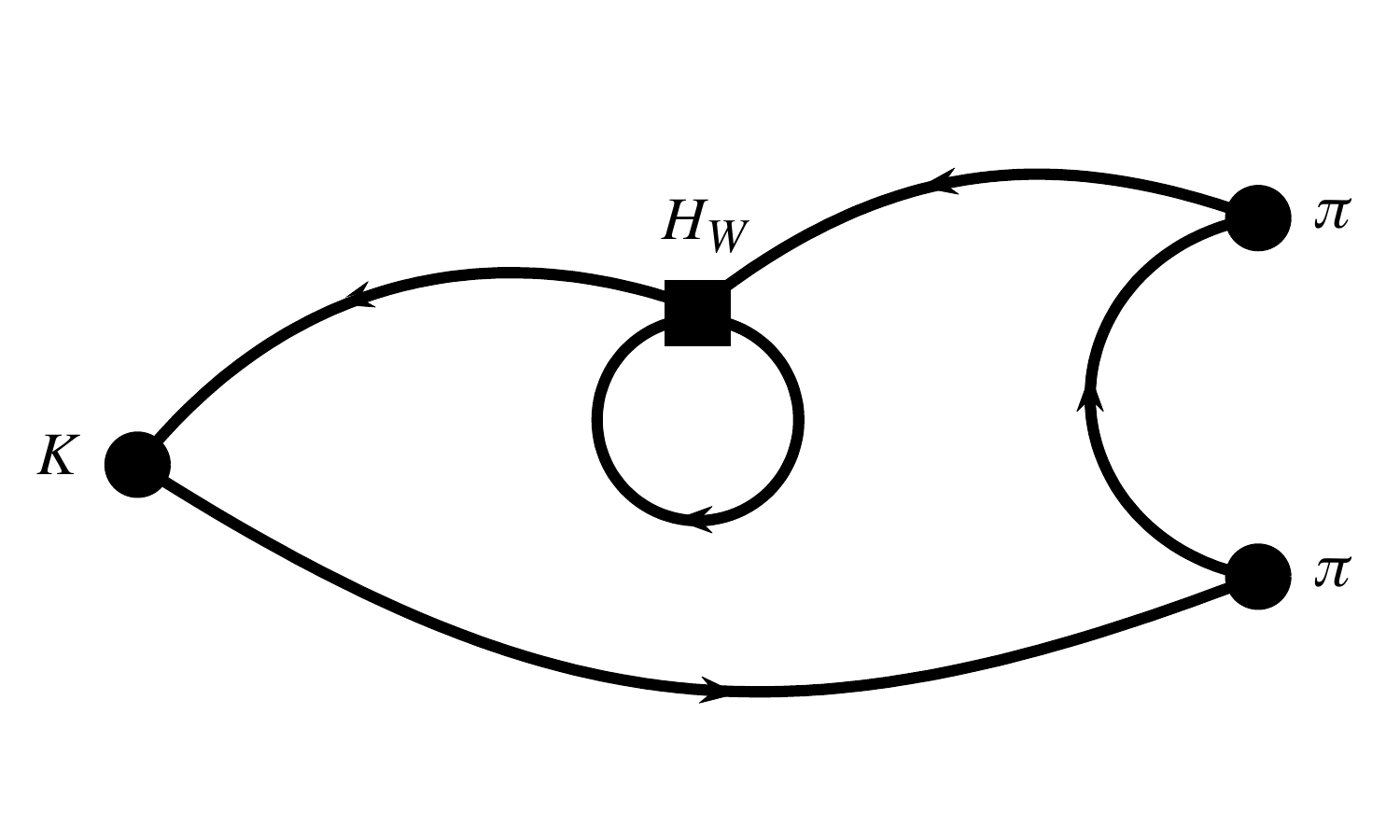}
\\
{\it type3}
\end{tabular}
\hspace{10mm}
\begin{tabular}{c}
\includegraphics[width=60mm]{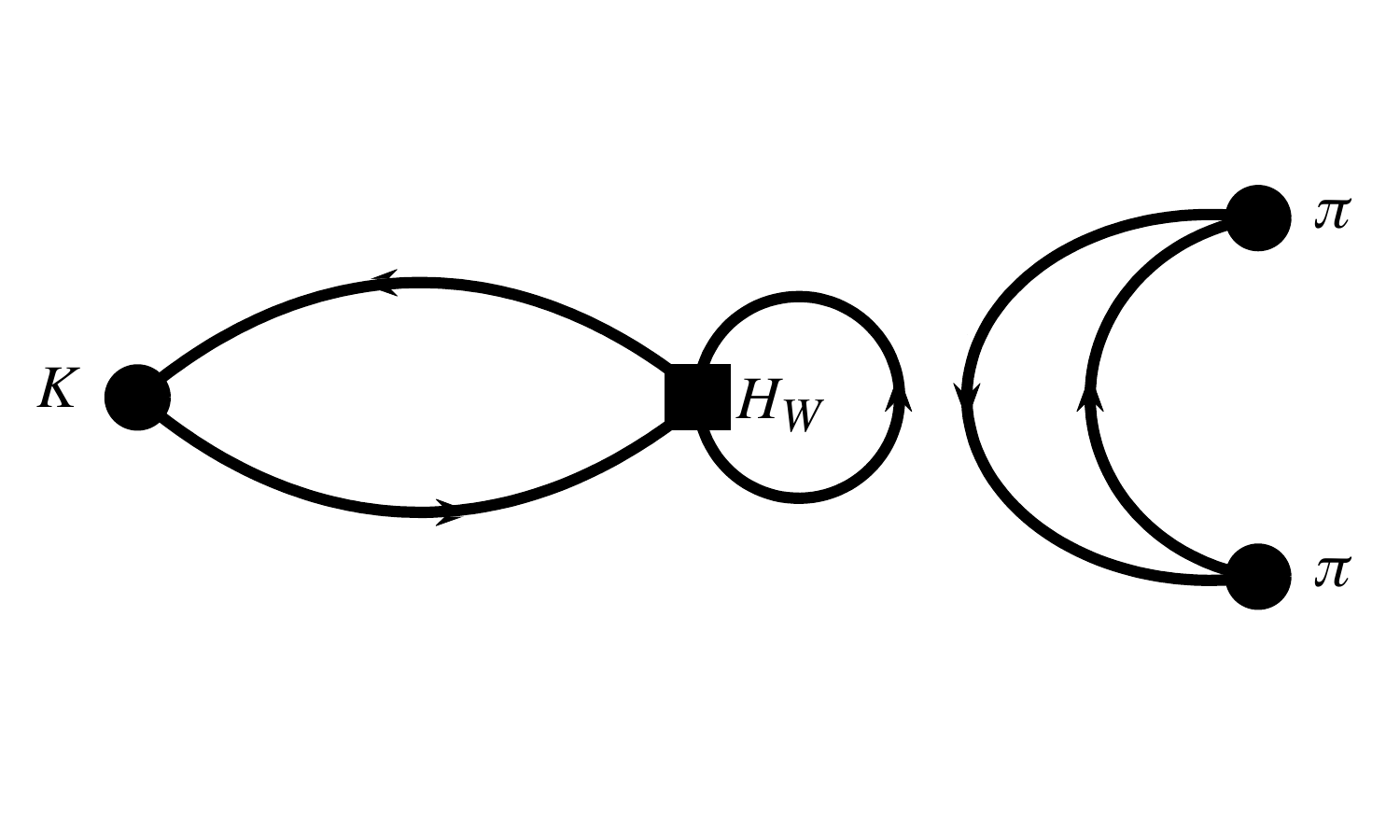}
\\
{\it type4}
\end{tabular}
\caption{
Diagrams that contribute to $K\to\pi\pi$ three-point functions.
}
\label{fig:diagrams}
\end{center}
\end{figure}

We calculate $K\to\pi\pi$ three-point functions
\begin{equation}
C^{\rm 3pt}_{\alpha,i}(t_1,t_2) = 
\left\langle O_{\pi\pi,\alpha}(t_{\pi\pi})^\dag Q_i(t_{op}) O_K(t_K) \right\rangle,\ \ \ \ 
t_1 = t_{\pi\pi} - t_{op}, t_2 = t_{op} - t_K,
\label{eq:3pt}
\end{equation}
where each operator is summed over spatial volume and $Q_i$ is
the $\Delta S=1$ four-quark operators~\cite{Buras:1993dy} relevant for $K\to\pi\pi$
decays.
Wick contractions of these correlation functions yield four typical
diagrams shown in Figure~\ref{fig:diagrams}.
We name these diagrams {\it type1} through {\it type4}.
While the $I=2$ channel only has the contribution from the {\it type1},
we need to calculate the four diagrams for the $I = 0$ channel.
For the $K\to\sigma$ correlation functions, there are the diagrams
that are analogous to the {\it type2--4} diagrams.

%

\begin{figure}[tbp]
\begin{center}
\begin{tabular}{c}
\includegraphics[width= 70mm]{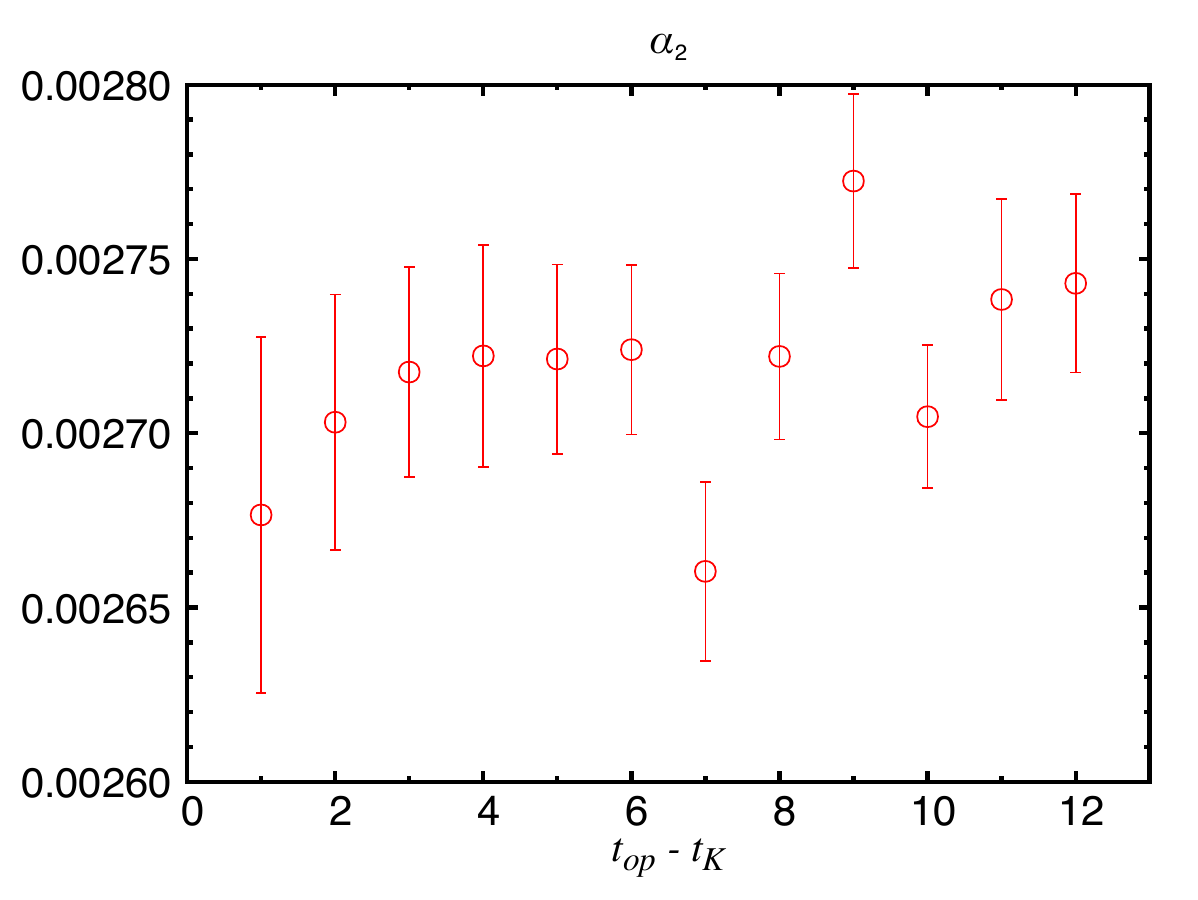}
\end{tabular}
\hfill
\begin{tabular}{c}
\includegraphics[width=70mm]{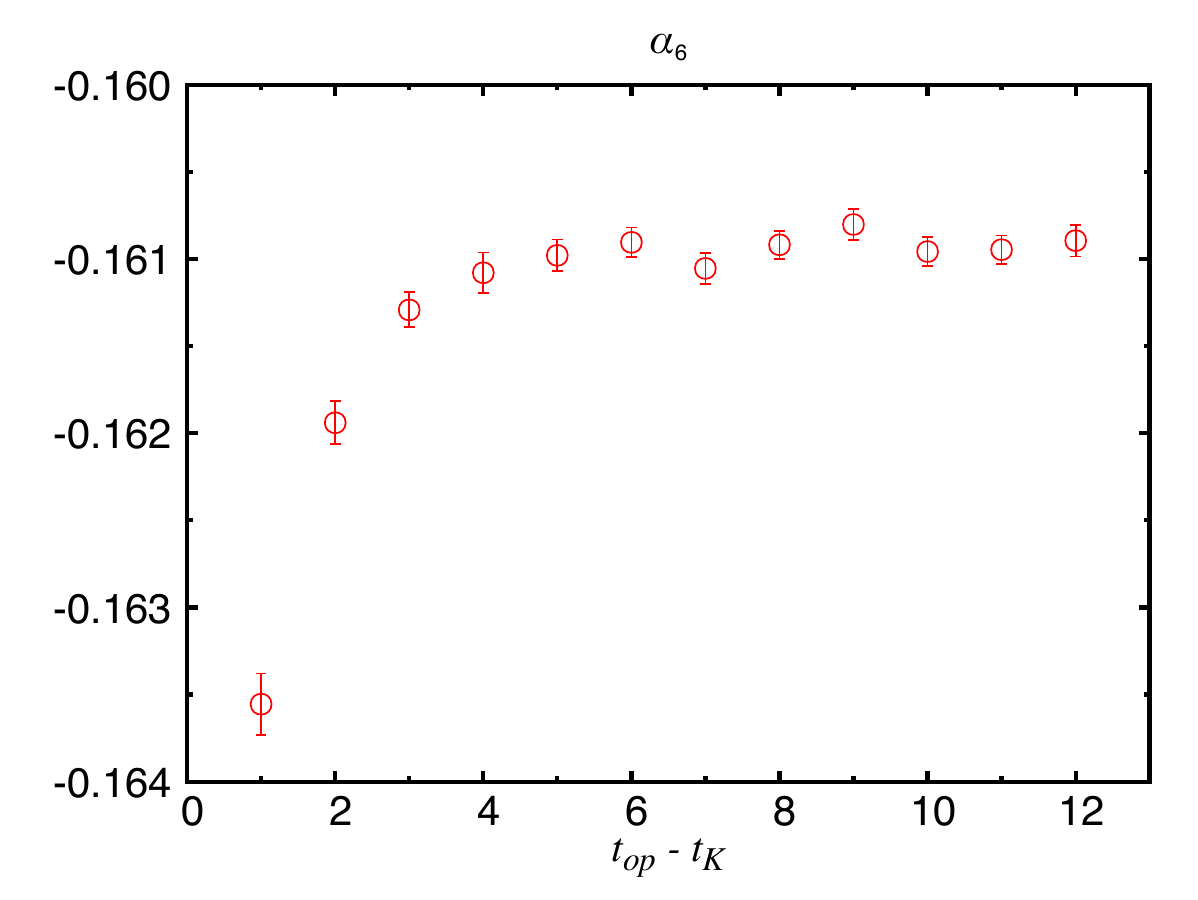}
\end{tabular}
\caption{
Subtraction coefficients $\alpha_2$ (left) and $\alpha_6$ (right) for $Q_2$ and $Q_6$, respectively.
}
\label{fig:subt}
\end{center}
\end{figure}

The presence of a quark loop in the {\it type3} and {\it type4} diagrams
induces a power divergence, which needs to be removed by
defining the subtracted four-quark operators 
\begin{equation}
Q_i = \hat Q_i - \alpha_i \bar s\gamma_5d,
\end{equation}
with the unsubtracted four-quark operators $\hat Q_i$.
We determine the subtraction coefficients $\alpha_i$ by imposing
the condition
\begin{equation}
\left\langle Q_i(t_{op}) O_K(t_K) \right\rangle = 0.
\label{eq:subt_cond}
\end{equation}
Figure~\ref{fig:subt} shows preliminary results for $\alpha_2$ (left)
and $\alpha_6$ (right).
We use these time-dependent subtraction coefficients for calculating the
three-point functions with the subtracted four-quark operators,
i.e.~we calculate the three-point functions at $t_2 = t_{op} - t_K$ using
the subtraction coefficients calculated at the same time separation $t_2 = t_{op}-t_K$.
This subtraction condition also removes the vacuum term 
$\langle O_{\pi\pi,\alpha}^\dag\rangle\langle Q_i(t_{op}) O_K(t_K) \rangle$
that could arise in the $K\to\pi\pi$ three-point functions~\eqref{eq:3pt} if
a different subtraction condition was imposed.
In what follows, the four-quark operators are understood as the subtracted ones.

The computational cost of the {\it type1} and {\it type2}
diagrams is as high as that of A2A propagators and much higher
than that of {\it type3} and {\it type4}.
Since the {\it type4} diagram, which is disconnected, is expected and
verified~\cite{RBC:2020kdj} to dominates the statistical error of the $I=0$ channel
of the three-point functions, 
the number of measurements of {\it type1} and {\it type2} diagrams can
be reduced without spoiling the statistical precision.
We perform the cost reduction by reducing the number of lattice sites of
the four-quark operators from $24^3$ to $8^3$ for each time slice.
With this reduction we can in principle take the contractions of {\it type1}
and {\it type2} diagrams $27\times$ faster.

\begin{figure}[tbp]
\begin{center}
\begin{tabular}{c}
\includegraphics[width=70mm]{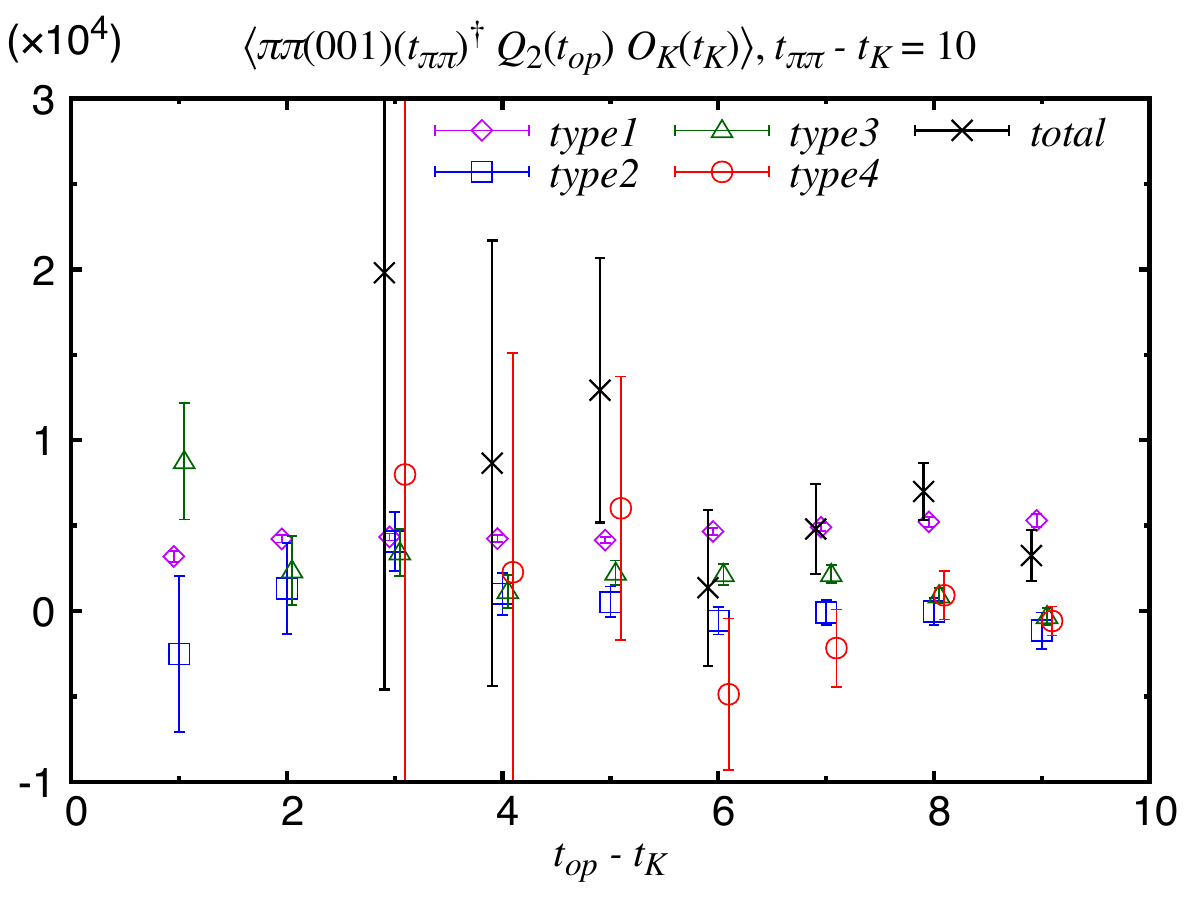}
\end{tabular}
\hfill
\begin{tabular}{c}
\includegraphics[width=70mm]{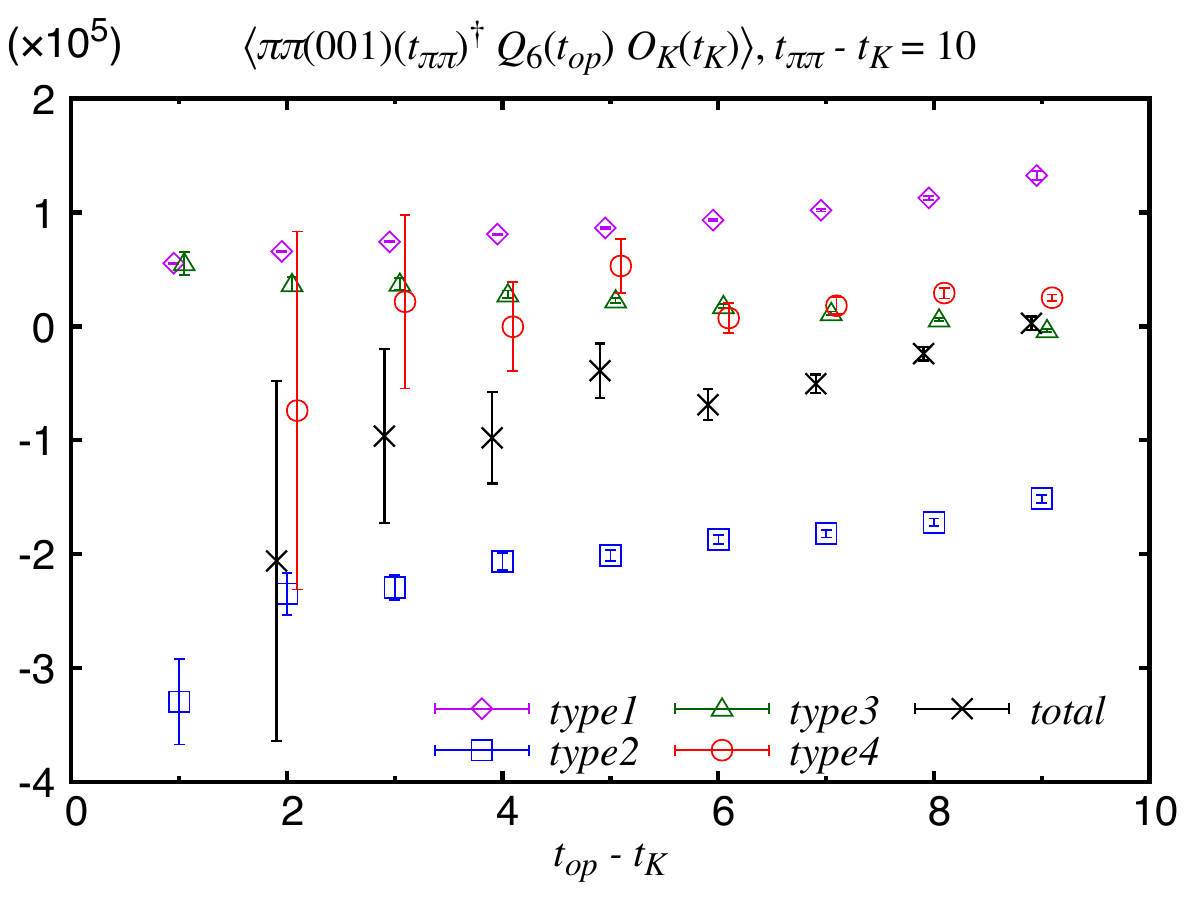}
\end{tabular}
\caption{
$K\to\pi\pi$ correlation functions with the four-quark operators $Q_2$ (left) and $Q_6$ (right)
and the $I=0$ two-pion operator $\pi\pi(001)$ at $t_{\pi\pi} - t_K = 10$ and their breakdowns
into the contributions of each diagram.
}
\label{fig:corr}
\end{center}
\end{figure}

Figure~\ref{fig:corr} shows the $K\to\pi\pi$ three-point functions with 
the four-quark operators $Q_2$ (left) and $Q_6$ (right) and the $I=0$ two-pion
operator $\pi\pi(001)$, which couples well with the first
excited two-pion state, which describes nearly on-shell kinematics in our
lattice setup.
The figure also shows the breakdown of the correlation functions
into the contributions of each diagram.
According to our $\pi\pi$ scattering study~\cite{Hoying:2020ghg},
the contaminations from higher excited states that are not taken
into account by the GEVP approach is significant at time separation
1 and 2 in lattice units and therefore we expect we need to extract
the $K\to\pi\pi$ matrix elements in the region $t_{\pi\pi} - t_{op} \ge 3$,
where the statistical error is  dominated by the {\it type4} diagram.
Thus the cost reduction of the {\it type1} and {\it type2}
diagrams does not appear to affect the statistical precision and
therefore is successful.

\section{Effective matrix elements}
\label{sec:efmlm}

Following the discussion in Ref.~\cite{Bulava:2011yz}, we calculate
the $K\to\pi\pi$ effective matrix elements 
\begin{equation}
M_{i,n}^{eff}(t_1,t_2) = 
\frac{\widetilde C^{\rm 3pt}_{n,i}(t_1,t_2)}
{\sqrt{\lambda_n^{t_1}{\rm e}^{-m_Kt_2}{\widetilde C^{\pi\pi}_n(t_1)} C^K(t_2)}},
\label{eq:efmlm}
\end{equation}
where we define state-projected correlation functions
\begin{align}
\widetilde C^{\rm 3pt}_{n,i}(t_1,t_2)
&= \sum_\alpha v_{n,\alpha}^* C_{\alpha,i}^{\rm 3pt}(t_1,t_2),
\\
\widetilde C^{\pi\pi}_n(t_1)
&= \sum_{\alpha,\beta} v_{n,\alpha}^* C_{\alpha\beta}^{\pi\pi}(t_1)v_{n,\beta},
\end{align}
with $\lambda_n$ and $v_{n,\alpha}$ being the eigenvalues and eigenvectors
extracted from a plateau region of the GEVP effective energies for the two-pion
correlation functions.

\begin{figure}[tbp]
\begin{center}
\begin{tabular}{c}
\includegraphics[width=70mm]{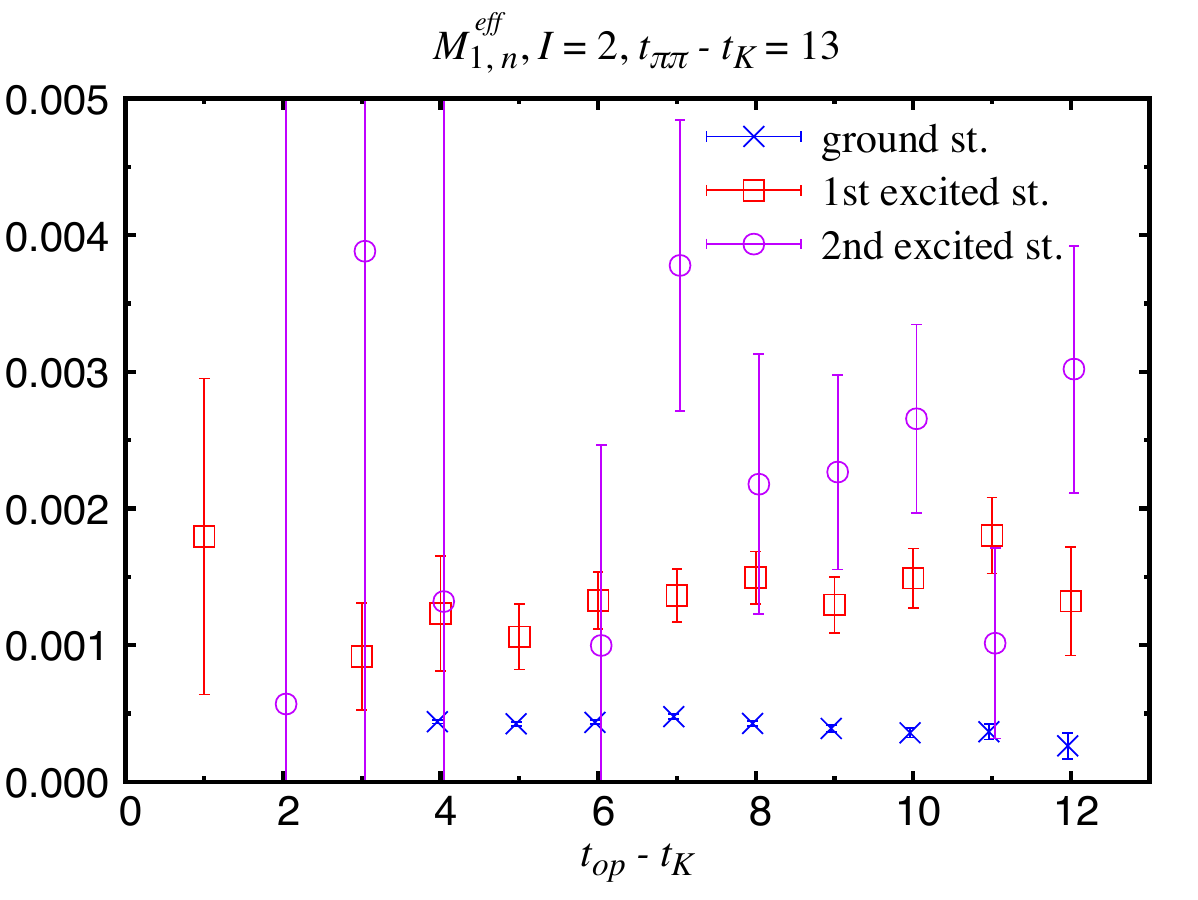}
\end{tabular}
\hfill
\begin{tabular}{c}
\includegraphics[width=70mm]{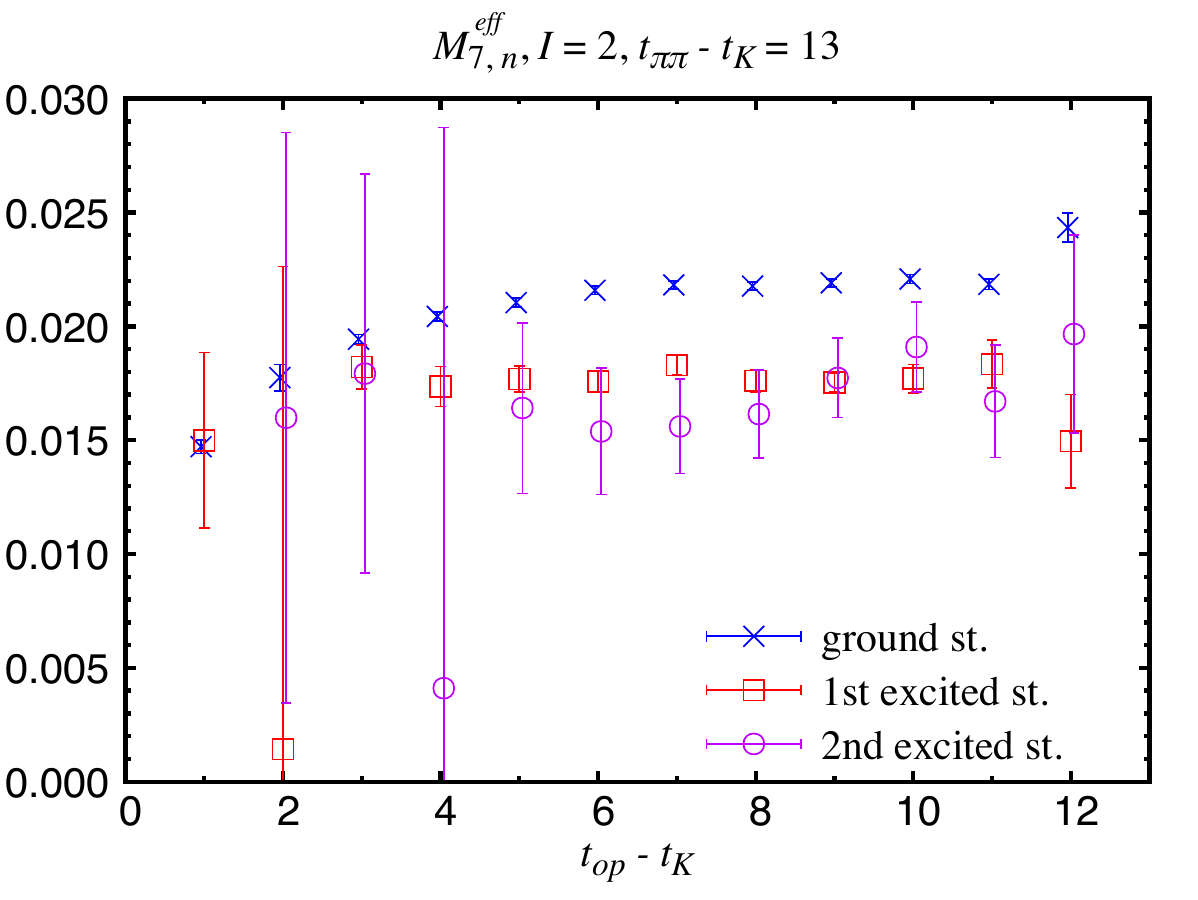}
\end{tabular}
\caption{
Preliminary results for $K\to\pi\pi$ effective matrix elements with the four-quark operators
$Q_1$ (left) and $Q_7$ (right) and the $I=2$ two-pion operators.
Here the two-pion operators $\pi\pi(000), \pi\pi(001), \pi\pi(011)$ and $\pi\pi(111)$ are used for GEVP.
}
\label{fig:efmlm2}
\end{center}
\end{figure}

Figure~\ref{fig:efmlm2} shows the $I=2$ channel of the effective
matrix elements with the four-quark operators $Q_1$ (left) and $Q_7$ (right).
We solve GEVP with the four two-pion interpolation operators
$\pi\pi(000),$ $\pi\pi(001),$ $\pi\pi(011)$ and $\pi\pi(111)$.
While the third excited two-pion state is also included in this GEVP analysis,
the corresponding matrix element is omitted from the plots since it has
a large statistical error.
We see a good plateau on the effective matrix elements with the
first excited two-pion state (squares), which describes nearly on-shell kinematics.

\begin{figure}[tbp]
\begin{center}
\begin{tabular}{c}
\includegraphics[width=70mm]{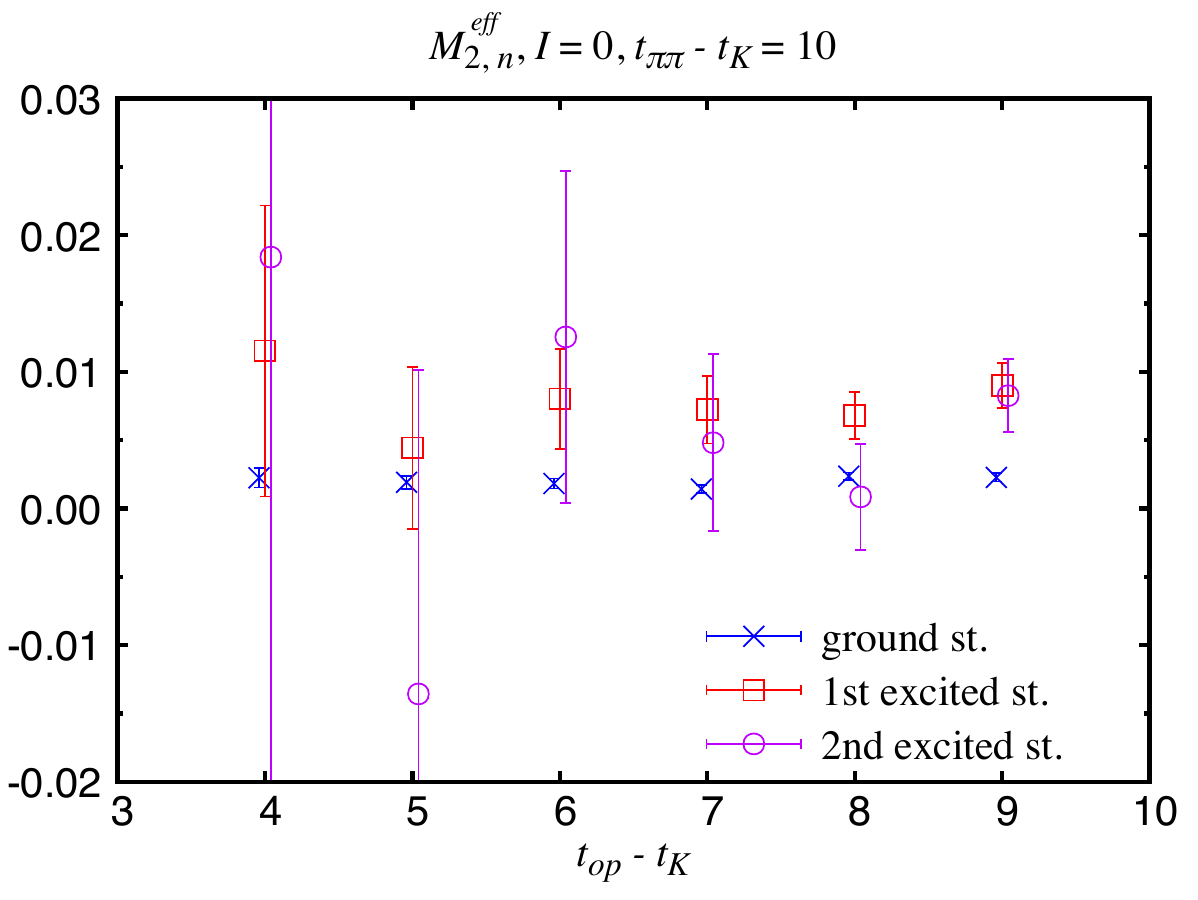}
\end{tabular}
\hfill
\begin{tabular}{c}
\includegraphics[width=70mm]{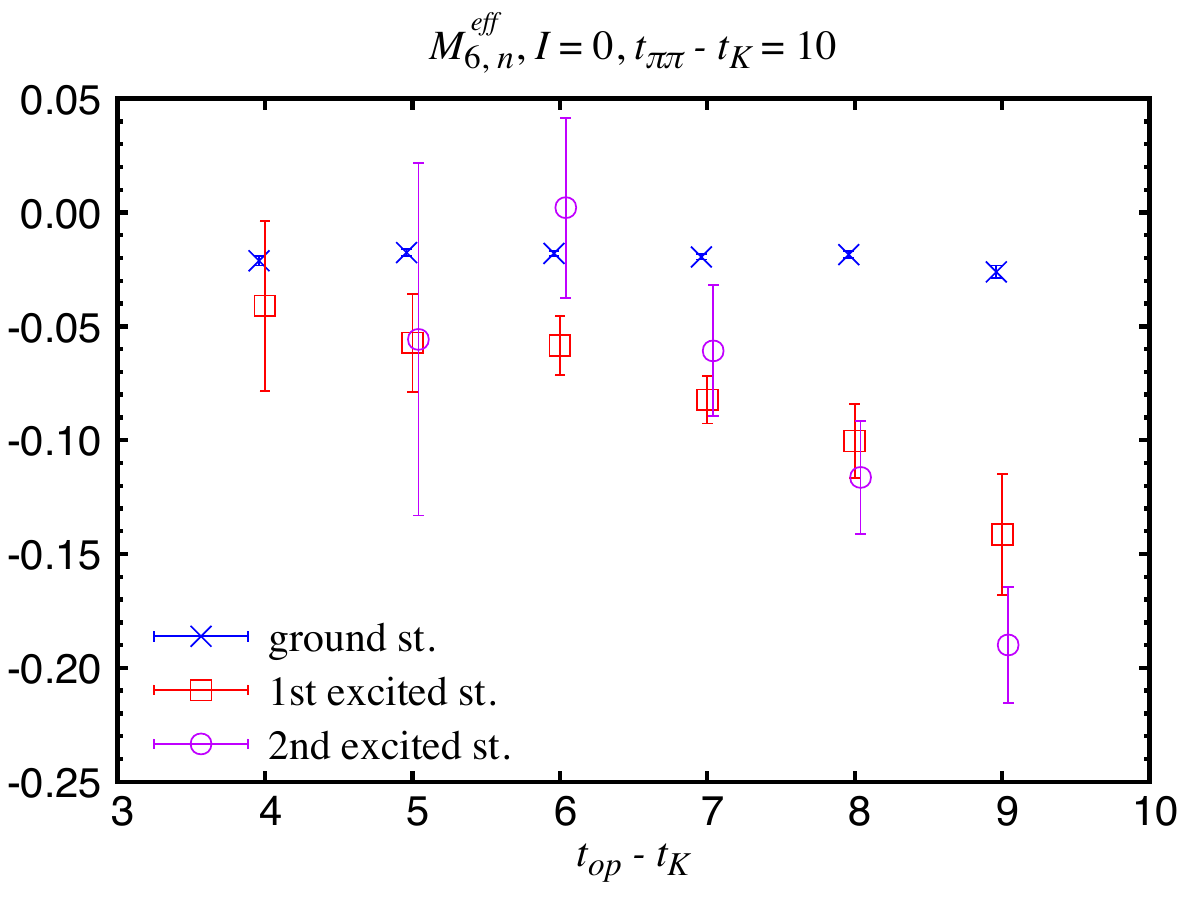}
\end{tabular}
\caption{
Preliminary results for $K\to\pi\pi$ effective matrix elements with the four-quark operators
$Q_2$ (left) and $Q_6$ (right) and the $I=0$ two-pion operators.
Here the three two-pion operators $\pi\pi(000), \pi\pi(001)$ and $\sigma$ are used for GEVP.
}
\label{fig:efmlm0}
\end{center}
\end{figure}

Figure~\ref{fig:efmlm0} shows the $I=0$ channel of the effective matrix elements
with the four-quark operators $Q_2$ (left) and $Q_6$ (right).
Here, the three two-pion interpolation operators $\pi\pi(000), \pi\pi(001)$ and
$\sigma$ are used for the GEVP analysis.
While the $I=0$ channel is noisier than the $I=2$ channel because of
the presence of a disconnected diagram
and we may need to increase the statistics and improve the analysis,
we still see some signal and plateau, which indicate the feasibility of calculating
the $I=0$ channel of $K\to\pi\pi$ matrix elements and $\varepsilon'$ with periodic
boundary conditions.

\section{Summary}
\label{sec:summary}

This study gives us a prospect that we could successfully calculate the
$K\to\pi\pi$ matrix elements and $\varepsilon'$ with periodic boundary
conditions.
We are attempting to decrease the statistical error by the factor of 3 in
order to reach the same precision as the one we had in our previous
calculation with G-parity boundary conditions~\cite{RBC:2020kdj}.
While we apply the AMA correction to the correlation functions of two-pion
operators, we have not introduced it to the $K\to\pi\pi$ three-point functions.
This makes the jackknife analysis nontrivial about noise cancelation 
due to the correlation between the numerator and denominator of
Eq.~\eqref{eq:efmlm}.
We are close to introduce the AMA correction to the three-point functions
and see how it improves the precision.

\acknowledgments{
M.T, T.B and L.C.J were supported in part by US DOE grant DE-SC0010339.  M.T. and L.C.J. were also supported in part by DOE Office of Science Early Career Award DE-SC0021147.  D.H was supported in part by a SCIDAC grant ``Computing the Properties of Matter with Leadership Computing Resources.''  T.I, C.J and A.S were supported in part by US DOE grant DE-SC0012704.  Computations for this work were carried out on facilities of the USQCD Collaboration, which are funded by the Office of Science of the U.S. Department of Energy.
}

\end{document}